\documentclass[10pt,conference]{IEEEtran}
\usepackage[left=0.7in, right=0.73in, top=0.7in, bottom = 0.98in]{geometry}
\usepackage{graphicx}
\usepackage{epstopdf}
\usepackage{setspace}
\usepackage{adjustbox} 
\DeclareGraphicsExtensions{.pdf,.jpeg,.png}
\usepackage{enumitem}
\usepackage[noadjust]{cite}
\usepackage{algorithm}
\usepackage{algorithmic}
\usepackage{hyperref}
\usepackage{subcaption}

\usepackage{amsmath}
\usepackage{lipsum}
\usepackage{mathtools}
\usepackage{cuted}
\usepackage{array}
\usepackage{amssymb}
\usepackage{caption}
\usepackage{multirow}
\usepackage{makecell}

\usepackage{xcolor}
\def\BibTeX{{\rm B\kern-.05em{\sc i\kern-.025em b}\kern-.08em
ThedeltaT\kern-.1667em\lower.7ex\hbox{E}\kern-.125emX}}

\begin{document}

\title{

Joint Beamforming and Duplex-Mode Segment Selection for Secure Uplink Transmission in Pinching Antenna Systems
}

\author{Zar Chi Phyo, Mohamed Elhattab, Chadi Assi, and Sanaa Sharafeddine  
}
\maketitle

\begin{abstract} 

Pinching antenna systems (PASS) have recently emerged as a promising architecture for reconfigurable, flexible and proximity-driven signal delivery. In this paper, we propose a segmented waveguide-based PASS architecture for enabling secure uplink communications, where each segment activates at most one pinching antenna (PA), which inherently eliminates the inter-antenna re-radiation effect and yields a tractable uplink signal model. To protect the uplink transmission against eavesdropping, the base station (BS) injects artificial noise (AN) toward the eavesdroppers via selected downlink segments using a duplex-mode segment selection mechanism, which dynamically configures each segment for either uplink reception or downlink AN transmission through switch-controlled fiber connections. We formulate the secrecy energy efficiency (SEE) maximization problem as a mixed-integer nonlinear program (MINLP) that jointly optimizes the receive beamforming vector, AN covariance, user transmit power, pinching antenna positions, and duplex-mode segment assignment. Given the non-convex and combinatorial nature of the problem, we develop an efficient alternating optimization (AO) framework, where the resulting subproblems are solved using the Dinkelbach method combined with successive convex approximation (SCA) and penalty-based relaxation. Simulation results demonstrate that the proposed framework significantly outperforms benchmark schemes in terms of SEE and robustness against waveguide attenuation. Moreover, AN effectively suppresses information leakage, while PA mobility, duplex-mode segment allocation, and waveguide segmentation further improve secrecy performance through flexible positioning and reduced in-waveguide propagation loss.

\end{abstract}

\begin{IEEEkeywords}
Physical layer security, pinching-antenna systems, , secrecy energy efficiency, segment selection, segmented waveguide, Uplink communication 
\end{IEEEkeywords}
\section{Introduction}


Future networks demand a fundamental rethinking of network design principles, with intelligence, energy efficiency, and autonomy at the core of network operation. In this context, artificial intelligence (AI)-driven control and agentic AI paradigms are expected to facilitate intelligent network reconfiguration and adaptive resource optimization under dynamic traffic conditions and heterogeneous service requirements. Beyond performance and efficiency, security remains an equally critical design imperative ~\cite{zhang20196g}. The inherent broadcast nature of wireless channels exposes transmissions to interception, eavesdropping, and information leakage threats that are further amplified in 6G environments characterized by an unprecedented scale of connected devices, diverse access technologies, and the pervasive exchange of sensitive personal, industrial, and mission-critical data. Conventional cryptographic approaches, while essential, increasingly face strain under the computational and latency constraints of resource-limited devices and real-time applications. This confluence of factors underscores the urgent need for scalable, adaptive, and computationally efficient security mechanisms that can operate natively within the intelligent and automated fabric of future 6G networks.


In this regard, physical-layer security (PLS) has emerged as a compelling alternative, exploiting the inherent randomness and variability of wireless channels to achieve information-theoretic confidentiality without sole reliance on computationally intensive encryption. Compared to traditional cryptographic schemes, PLS offers a lightweight security mechanism with significantly reduced computational overhead and latency. 
Among PLS techniques, artificial noise (AN) injection has been widely adopted as an effective strategy to deliberately degrade the eavesdropper's reception while preserving the quality of the intended link. However, the continuous transmission of AN inevitably increases power consumption, introducing a fundamental tradeoff between achievable secrecy rate and energy efficiency. This motivates the adoption of secrecy energy efficiency (SEE) as a new unified performance metric capturing both secure transmission capability and energy awareness within a single design objective~\cite{xu2025secure,bao2023secrecy}. As 6G networks are expected to simultaneously deliver robust security and stringent energy efficiency targets, SEE emerges as a particularly relevant and practically meaningful criterion for the design of next-generation secure wireless systems.

Realizing the full potential of PLS in practice, however, requires not only sophisticated signal processing strategies but also the ability to actively shape and control the wireless propagation environment. Toward this end, reconfigurable antenna technologies have attracted considerable research interest as enablers of dynamic channel reconfiguration, including reconfigurable intelligent surfaces (RIS), movable antennas, and fluid antennas~\cite{wong2020fluid,zhu2023modeling}. These technologies enhance PLS performance by dynamically reshaping the propagation environment to strengthen the legitimate channel while simultaneously suppressing the eavesdropper's reception. Despite their respective merits, these approaches are often constrained by limited spatial flexibility, restricted coverage range, or practical deployment complexity, which may limit their effectiveness in highly dynamic and heterogeneous 6G scenarios. To overcome these limitations, pinching-antenna systems (PASS) have recently emerged as a promising flexible-antenna architecture, wherein signals are guided along low-loss dielectric waveguides and radiated at arbitrary desired locations via selectively activated pinching antennas (PAs), enabling meter-scale spatial reconfigurability and enhanced line-of-sight (LoS) communication~\cite{suzuki2022pinching,liu2025pinching,yang2025pinching,ding2025flexible}. This unprecedented degree of spatial controllability introduces new and powerful degrees of freedom for secure communication design, positioning PASS as a particularly attractive candidate for PLS in next-generation wireless networks. 

Nevertheless, conventional PASS architectures face a significant challenge in uplink communications stemming from the inter-antenna re-radiation (IAR) effect, which arises when multiple PAs are co-located along a single waveguide and induces complex mutual coupling that renders the resulting signal model computationally intractable~\cite{hou2025performance,bereyhi2025mimo}. To overcome this fundamental bottleneck, segmented waveguide-enabled pinching-antenna systems (SWANs) have recently been proposed, where the long waveguide is partitioned into multiple short, electrically isolated segments each equipped with an independent feed point~\cite{ouyang2026uplink}. By restricting activation to a single PA per segment, SWAN inherently eliminates the IAR effect and yields a tractable closed-form uplink signal model, while simultaneously reducing in-waveguide propagation loss and improving system maintainability and modularity. Collectively, these attributes make SWAN a compelling and practically viable platform for the flexible, efficient, and secure design of uplink systems in next-generation wireless networks, motivating its adoption as the architectural foundation of this work.


\subsection{Prior Works}
Building on the above discussion, significant research has been conducted on conventional PASS to exploit their flexible antenna positioning capability offered by dielectric waveguides ~\cite{suzuki2022pinching,liu2025pinching,yang2025pinching,ding2025flexible}. Early studies mainly focused on downlink transmission, where antenna activation, resource allocation, and placement design were investigated for single- and multi-waveguide architectures under various multiple access schemes ~\cite{10912473,wang2025antenna,xie2025low}. More recently, PASS has also been integrated with coordinated multipoint (CoMP) transmission, where the achievable sum rate was maximized through the joint optimization of transmit beamforming and pinching antenna locations under users' QoS requirements ~\cite{amhaz2025coordinated}. In particular, downlink rate and sum-rate maximization problems have been extensively studied using Non-Orthogonal Multiple Access (NOMA) and Rate-Splitting Multiple Access (RSMA) in ~\cite{10896748,zhou2025sum,wang2025sum}, while uplink PASS remains comparatively underexplored despite some initial works on rate optimization and multiuser scenarios ~\cite{tegos2025minimum,hou2025performance,zhang2025uplink,bereyhi2025mimo}. A joint uplink-downlink PASS framework was recently proposed in \cite{khisa2025joint}, where antenna activation and resource allocation were jointly optimized to maximize the overall system sum rate. The authors in~\cite{phyo2025joint} investigated uplink RSMA-assisted PASS and maximized the achievable sum rate through the joint optimization of user transmit power and continuous antenna positioning using an AO-SCA framework.

Recent studies have extended PASS to SWAN architectures, enabling more flexible uplink system designs. Existing studies have conducted SWAN architectures for both uplink and downlink systems. In~\cite{ouyang2026uplink}, different protocols for SWAN were studied, demonstrating improved flexibility and performance. The tri-hybrid uplink beamforming, which jointly optimizes digital, analog, and pinching beamforming, was investigated in~\cite{jiang2026exploiting}, while the maximization of uplink sum-rate was studied in~\cite{gu2025sum}. The framework was further extended to integrated sensing and communication (ISAC) systems in~\cite{jiang2025segmented}, and the computation--communication tradeoff was analyzed in~\cite{gan2025revealing}. These studies consistently show the performance advantages of segmentation, but primarily focus on rate-oriented metrics under non-secure communication systems.

From a security and energy-efficiency (EE) perspective, SEE has emerged as an important performance metric that jointly measures security rate and power consumption~\cite{zappone2016energy}. In this context, advanced reconfigurable antenna architectures, particularly PASS, have attracted significant attention due to their ability to dynamically adjust antenna locations along dielectric waveguides, allowing flexible beamforming with a limited number of radio frequency (RF) chains~\cite{suzuki2022pinching,liu2025pinching,yang2025pinching,ding2025flexible}. 

In parallel, physical layer security for PASS has attracted increasing attention, mainly in downlink scenarios. In~\cite{wang2025pinching}, the secrecy rate maximization for PASS was studied using antenna activation and phase alignment techniques. The work in~\cite{zhu2025pinching} extended this framework to both single and multi-waveguide scenarios, incorporating waveguide division and multiplexing strategies with artificial noise (AN) to enhance secrecy performance. Multi-user and multi-eavesdropper scenarios were considered in~\cite{papanikolaou2025secrecy}, where AN-aided transmission was optimized for secrecy rate maximization. In~\cite{musri2025adaptive}, a meta-learning-based approach was proposed to adapt antenna positioning for secrecy enhancement in dynamic settings, while~\cite{sun2025physical,li2025secrecy} further investigated secrecy performance under different PASS configurations. Despite these advances, existing studies focus mostly on downlink transmission, where introducing artificial noise and antenna management are centered at the base station, and do not address uplink secure communication.

From an EE perspective, several works have investigated EE optimization for PASS, considering factors such as waveguide propagation loss, RF chain consumption, and system power models. In~\cite{zeng2025energy}, EE maximization with quality-of-service (QoS) constraints was studied for downlink PASS systems. The authors in~\cite{zeng2025energyNOMA,zeng2025sum} considered energy-efficient resource allocation for uplink PASS under NOMA transmission, while~\cite{zhou2025spectral} examined the trade-off between spectral efficiency and energy efficiency using multi-objective optimization techniques. Although these studies show the potential of PASS to improve the network EE, they consider separate objectives for EE and security.

In a wider context, SEE has been extensively studied in other reconfigurable wireless architectures. For instance, SEE maximization in RIS and AN-assisted Multiple-input single- output (MISO) systems was investigated in~\cite{xu2025secure}, while secure communication in movable-antenna-enabled full-duplex systems was studied in~\cite{guo2024movable,ding2024secure}. In RIS-assisted networks, SEE optimization has been considered for both active and nearly-passive RIS architectures in~\cite{fotock2025secrecy}, as well as for distributed RIS-assisted uplink systems in~\cite{bao2023secrecy}. These works provide important insights into joint security and energy-efficient design; however, they are not directly applicable to PASS due to fundamental differences in channel modeling and antenna control mechanisms.


\subsection{Motivation and Contributions}

Despite the growing body of research on PASS-based communications, segmented waveguide architectures, physical layer security, and  energy-efficient transmission, the intersection of these paradigms remains largely unexplored. Specifically, and to the best of our knowledge, no existing work has addressed secure uplink communication in segmented waveguide-enabled PASS under an artificial noise-aided design with a SEE objective. Current contributions either focus on non-secure performance metrics such as spectral efficiency and sum rate, or address downlink secrecy scenarios, leaving the uplink secure transmission problem, which introduces fundamentally different signal modeling challenges due to the inter-antenna re-radiation effect, unaddressed. This gap motivates the present work.

The design of secure uplink SWAN systems introduces several new challenges.  First, The segmented architecture enables each waveguide segment to be independently configured for either uplink reception or downlink AN transmission. This duplex-mode flexibility, while beneficial for system adaptability, results in a tightly coupled mixed-integer optimization problem in which discrete segment assignment and continuous beamforming variables are deeply intertwined. Second, although adding AN effectively degrades the eavesdropper's reception, it inevitably increases the total power consumption. This introduces a fundamental tradeoff between secrecy performance and energy efficiency which cannot be captured by the secrecy rate alone, motivating the adoption of SEE as the primary design objective. Finally, the resulting SEE maximization problem is highly non-convex due to the fractional nature of the objective function, the non-linear coupling among transmit power, receive beamforming, AN design, antenna positioning, and segment selection variables, and the presence of binary assignment constraints, rendering standard convex optimization techniques directly inapplicable.

Motivated by the above challenges, we propose an AN-aided secure uplink communication framework, with the objective of maximizing SEE through joint optimization of all system design variables. The main contributions are summarized as follows:

\begin{itemize}
    \item We propose an AN-aided secure uplink framework for segmented PASS, in which each waveguide segment is dynamically and independently configured for either uplink reception or downlink AN transmission via a duplex-mode selection mechanism controlled by switch-enabled fiber connections. This design eliminates the inter-antenna re-radiation effect while enabling flexible and hardware-efficient cooperative security.

    \item We formulate a maximization problem that jointly optimizes the receive beamforming vector, AN covariance matrix, user transmit power, pinching antenna positions, and duplex-mode segment selection, subject to QoS in terms of the minimum rate requirements, power budget at the users, and segment-selection constraints. The resulting problem is a mixed-integer nonlinear program (MINLP) which captures the fundamental tradeoff between secrecy performance and energy consumption in SWAN systems.

    \item We develop a computationally efficient optimization framework that integrates the Dinkelbach method, alternating optimization (AO), successive convex approximation (SCA), and penalty-based relaxation to address the non-convex fractional objective and the mixed discrete-continuous variable structure. \textit{The convergence of the proposed algorithm is established} and its computational complexity is analyzed.

    \item Simulation results validate the effectiveness of the proposed framework and demonstrate its superiority over benchmark schemes. In particular, we reveal the critical role of duplex-mode segment allocation in governing the secrecy-energy tradeoff, and provide actionable design guidelines for the deployment of secure uplink SWAN systems in next-generation wireless networks.
\end{itemize}

\subsection{Organization and Notations}

The remainder of this paper is organized as follows. 
Section~\ref{sec:system} introduces the system model. 
Section~\ref{sec:problem} formulates the SEE maximization problem. 
Section~\ref{sec:algorithm} presents the proposed solution framework. 
Section~\ref{sec:results} provides numerical results. 
Finally, Section~\ref{sec:conclusion} concludes the paper.

\textit{Notations}: Scalars, vectors, and matrices are denoted by lower-case, boldface lower-case, and boldface upper-case letters, respectively. 
$\mathbb{C}^{M \times 1}$ and $\mathbb{R}^{M \times 1}$ denote the spaces of complex- and real-valued matrices. 
$(\cdot)^T$, and $(\cdot)^H$, and $(\cdot)^*$ represent transpose, conjugate transpose, and complex conjugate, respectively. 
$\|\cdot\|$ and $|\cdot|$ denote the Euclidean norm and absolute value, respectively. 
$\mathbb{E}\{\cdot\}$ denotes the expectation operator. 
For a vector $\mathbf{x}$, $[\mathbf{x}]_m$ denotes its $m$-th element.
The notation $\mathcal{CN}\mathbf{0},\sigma_B^2\mathbf{I})$ refers to the complex Gaussian distribution with mean 0 and variance $\sigma_B^2$.
The notation $\mathbf{I}_M$ denotes the $M \times M$ identity matrix.
The notation $\nabla_{\boldsymbol{x}}$ denotes for the gradient information with respect to each respective variables.
The operator $[x]^+ \triangleq \max(x,0)$ denotes the positive part operator. 
The symbol $j=\sqrt{-1}$ denotes the imaginary unit. Finally, $\mathcal{O}(.)$ denotes big-O notation.

\section{System Model}
\label{sec:system}
\begin{figure}
    \centering
\includegraphics[width=0.9\columnwidth]{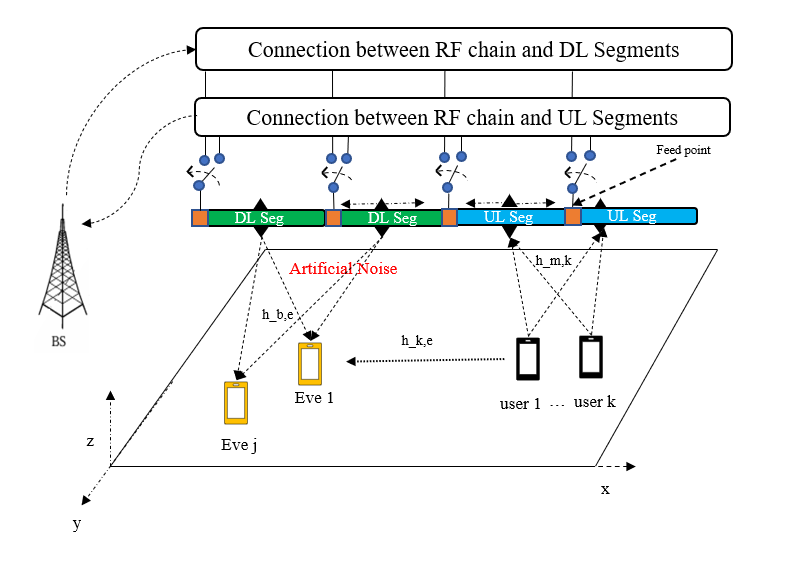}
    \caption{Illustration of a secure uplink communication system}
    \label{fig:Illustration of secure uplink communication system}
\end{figure}

As illustrated in Fig. \ref{fig:Illustration of secure uplink communication system}, we consider a secure uplink communication system where a PASS-based full-duplex base station (BS) serves multiple uplink users in the presence of an eavesdropper, denoted as Eve. The BS employs a segmented PASS \cite{ouyang2026uplink} consisting of $M$ waveguide segments to receive signals from uplink users via uplink segments (UL seg) while simultaneously transmitting AN via downlink segments (DL seg) to interfere with the eavesdropper. These segments are physically independent yet arranged sequentially along the x-axis. The users and eves are distributed in a rectangular service region with dimensions $D_x$ and $D_y$ along the $x$- and $y$-axes, respectively. The segmented waveguide is deployed parallel to the $x$-axis at a height $d$ above the user plane. The set of segments is denoted by $\mathcal{M} = \{1,2,...,M\}$ arranged end-to-end, where each segment has a length $L$, such that $D_x = LM$. The position of the $k$-th uplink user equipment (UE) is denoted by $\psi_k = (x_k,y_k,0)$ where $k \in \mathcal{K} = {1, 2, \dots, K}$. Let $\psi_0^m = (\psi_0^m,y_m,d)$ denote the location of the feed point of the $m$-th segment, and the feed point of each segment is placed at the front-left end of the segment for consistency. Each segment consists of one PA, which can be configured with each segment~\cite{jiang2025segmented}. The location of the PA in the $m$-th segment is denoted by $\psi_m = (x_m,y_m,d)$. The position of the eavesdropper is denoted by $\psi_{j}^e = (x_{j}^e,y_{j}^e,0)$ where $j \in \mathcal{J} =\{1,2, \dots, J\} $, and $\mathcal{J}$ is the set of Eves.

\subsection{Channel Modeling}
The free-space channels between the $m$-th uplink segment, $\forall m \in \mathcal{M}$, and the $k$-th user, and between the $m$-th downlink segment, $\forall m \in \mathcal{M}$ and the eavesdropper $j$-th Eve, $\forall j \in \mathcal{J}$, are respectively given by
\begin{subequations}\label{eq:free_space_channels}
\begin{align}
h_{m,k}^{r} &=
\frac{\sqrt{\eta}\, e^{-j\frac{2\pi}{\lambda}\|\psi_k-\psi_m^{r}\|}}
{\|\psi_k-\psi_m^{r}\|}
\label{eq:free_space_channels_a}, \\
h_{m,j}^{t} &=
\frac{\sqrt{\eta}\, e^{-j\frac{2\pi}{\lambda}\|\psi_e-\psi_m^{t}\|}}
{\|\psi_{j}^e-\psi_m^{t}\|},
\label{eq:free_space_channels_b}
\end{align}
\end{subequations}
where $\eta=\frac{c^2}{16\pi^2 f_c^2}$, $c$ is the speed of light, $f_c$ represents the carrier frequency, and $\lambda$ is the wavelength. $\|\psi_k-\psi_m^r\|$ represents the Euclidean distance between UE-$k$ and PA at waveguide $m$ and $\|\psi_{j}^e-\psi_m^t\|$ denotes the distance between PA and the Eve.\\
The in-waveguide channel between the feed point and the $m$-th PA segment for the receive and transmit modes are respectively expressed as
\begin{subequations}\label{eq:waveguide_channel}
\begin{align}
g_m^{r} &=
10^{-\frac{\kappa}{20}\|\psi_m^{r}-\psi_{0}\|}
e^{-j\frac{2\pi}{\lambda_g}\|\psi_m^{r}-\psi_{0}\|},
\label{eq:waveguide_channel_a} \\
g_m^{t} &=
10^{-\frac{\kappa}{20}\|\psi_m^{t}-\psi_{0}\|}
e^{-j\frac{2\pi}{\lambda_g}\|\psi_m^{t}-\psi_{0}\|},
\label{eq:waveguide_channel_b}
\end{align}
\end{subequations}
where $\kappa$ denotes the in-waveguide loss measured in dB/m, and $\lambda_{g}=\frac{\lambda}{\eta_{eff}}$ represents the waveguide wavelength and $\eta_{eff}$ represents the effective refractive index of the waveguide. $\|\psi_m^r-\psi_{0}^m\|$ and $\|\psi_m^t-\psi_{0}^m\|$ denote the Euclidean distances from the $m$-th PA to the feed points of the receiving ($r$) and transmitting ($t$) segments, respectively \cite{wang2025pinching}. 

\subsection{Uplink Transmission Model}
The uplink channel can be expressed as:
\begin{equation}
\mathbf{h}_k =
\big[
\alpha_1{h}_{1,k}^r g_1^r,\;
\alpha_2{h}_{2,k}^r g_2^r,\;
\ldots,\;
\alpha_M{h}_{M,k}^r g_M^r
\big]^T \in \mathbb{C}^{M \times 1},
\label{e3}
\end{equation}
where $\alpha_m=\{0,1\}$ is the uplink segment selection variable ($\alpha_m=1$, the $m$-th segment is activated for uplink reception and $\alpha_m=0$, otherwise).

Let $\mathcal{M}_{\rm{UL}}=\{m \mid \alpha_m=1\}$ denote the set of selected uplink segments, where the total number of active segments is ${M}_{UL}=|\mathcal{M}_{UL}|$ \cite{khisa2025joint}. During the uplink transmission phase, the users transmit their signals to the segmented PA receiver. Accordingly, the received signal from user $k$, $\forall k \in \mathcal{K}$, at the BS can be expressed as follows:
\begin{equation}
\footnotesize
\begin{aligned}
y_{k} =
\frac{1}{\sqrt{M_{UL}}}
\left(
\mathbf{w}_k^H \mathbf{h}_k \sqrt{p_k}s_k
+
\sum_{i \neq k}^{K}
\mathbf{w}_k^H \mathbf{h}_i \sqrt{p_i}s_i
+
\mathbf{w}_k^H \mathbf{n}_{BS}
\right),
\end{aligned}
\label{eq:received_signal}
\end{equation}
where $\mathbf{w}_k \in \mathbb{C}^{M \times 1}$ is the receive beamformer, $p_k$ is the user transmit power, and where $\mathbf{n}_{BS} \sim \mathcal{CN}( \mathbf{0},\sigma_B^2\mathbf{I})$ denotes the additive white Gaussian noise (AWGN) vector at the BS, and $s_k$ is the data symbol of user $k$ with $\mathbb{E}[|s_k|^2]=1$. 

\subsection{Downlink Transmission Model}
We assume the BS generates an AN signal and transmits it through the selected downlink segments towards Eve, to protect the leaked information from uplink transmissions. Let $\mathbf{z} \in \mathbb{C}^{M \times 1}$ denote the transmitted AN signal, which follows $\mathbf{z} \sim \mathcal{CN}(\mathbf{0},\mathbf{V})$ where $\mathbf{V}=\mathbb{E}\{\mathbf{z}\mathbf{z}^{H}\}\succeq \mathbf{0}$ is the AN covariance matrix \cite{papanikolaou2025secrecy}. Furthermore, the downlink channel between the BS and the eavesdropper is given by \cite{wang2025pinching} 
\begin{equation}
\textbf{h}_{b,j} =[\beta_1 h_{1,j}^t{g}_1^t,...,\beta_M h_{M,j}^t{g}_M^t ]^T \in \mathcal{C}^ {M \times 1},
\end{equation}
where $\beta_m=\{0,1\}$ denotes the downlink segment selection variable. 
Let $\mathcal{M}_{DL}=\{m \mid \beta_m=1\}$ denote the set of selected downlink segments, where the total number of active downlink segments is given by $M_{DL}=|\mathcal{M}_{DL}|$ \cite{khisa2025joint}. Therefore, the signal received at Eve-$j$ can be expressed as:
\begin{equation}
y_{j}^e = \frac{1}{\sqrt{{{M}_{DL}}}}\mathbf{h}_{b,j}^H\mathbf{z}+  \sum_{k=1}^Kh_{k,j}\sqrt{p_k}s_k+ n_e ,
\label{ye}
\end{equation}
where $h_{k,j}$ is the channel between user-$k$ and Eve-$j$.
\begin{equation}    h_{k,j}=\frac{{\sqrt{\eta}} e^{-j\frac{2\pi}{\lambda}\|\psi_k-\psi_e\|}}{\|\psi_k-\psi_{j}^e\|}.
    \label{e1}
\end{equation}

\subsection{Achievable Rate Analysis}
According to \eqref{eq:received_signal} and \eqref{ye}, the received signal-to-interference-plus-noise ratio (SINR) of UE-$k$ at the BS can be expressed as

\begin{equation}
\gamma_k=
\frac{p_k |\mathbf{h}_k^H \mathbf{w}_k|^2}
{\sum_{i\neq k} p_i |\mathbf{h}_i^H \mathbf{w}_i|^2
+\|\mathbf{w}_k\|^2 \sigma_k^2 } ,
\label{eq:SINR_K}
\end{equation}
where the numerator denotes the desired signal power of UE-$k$, while the first term in the denominator represents the inter-user interference from other uplink users and the second term corresponds to the noise power at the BS.
Similarly, the SINR of the $j$-th eavesdropper is given by

\begin{equation}
\gamma_{j,k}=
\frac{p_k |h_{k,j}|^2}
{\sum_{i\neq k} p_i |h_{i,j}|^2
+\frac{1}{M_{DL}}\mathbf{h}_{b,j}^H
\mathbf{V}\mathbf{h}_{b,j}
+\sigma_e^2},
\label{eq:SINR_E}
\end{equation}
where the denominator consists of the multiuser interference, AN injected by the downlink segments, and the noise power at the Eve.
Accordingly, the achievable rates of UE-$k$ and Eve-$j$ are respectively given by

\begin{equation}
R_k=\log_2(1+\gamma_k),
\label{Rk_c}
\end{equation}

\begin{equation}
R_{j,k}=\log_2(1+\gamma_{j,k}).
\label{Rk_e}
\end{equation}
Since the secrecy rate is limited by the most powerful eavesdropper and $R_{sec}$ cannot be less than 0, the achievable secrecy rate can be defined as 
\begin{equation}
\begin{aligned}
R^{\mathrm{sec}} 
&= \sum_{k=1}^{K} 
\left[ 
R_k - \max_{j \in \mathcal{J}} R_{j,k} 
\right]^+ \\
&= \sum_{k=1}^{K} 
\max \left\{ 
R_k - \max_{j \in \mathcal{J}} R_{j,k}, \; 0 
\right\},
\end{aligned}
\label{eq:R_sec}
\end{equation}
On the other hand, the total power consumption of the system can be given by
\begin{equation}
    P_{tot}=\xi \left(\mathrm{Tr}(\mathbf V)+\sum_{k=1}^Kp_k\right)+P_c ,
\end{equation}
where $P_c$ is the circuit power consumption and $\xi \in (0,1]$ is the inverse of the power amplifier’s efficacy at the transmitter \cite{zeng2025energy}. Therefore, the secrecy energy efficiency of the system is defined as:

\begin{equation}
    \text{SEE} = \frac{R^{sec}}{P_{tot}}.
\end{equation}

\section{Problem Formulation}\label{sec:problem}
We aim to maximize the SEE of the uplink UEs while guaranteeing the minimum rate requirements by jointly optimizing the uplink transmit power vector $\mathbf{p}$, receive beamforming vectors $\mathbf{w}_k$, artificial noise covariance matrix $\mathbf{V}$, antenna position vector $\mathbf{x}^p$, and uplink/downlink segment selection variables $\boldsymbol{\alpha}$ and $\boldsymbol{\beta}$, respectively, as follows.
\allowdisplaybreaks
\begin{subequations}
\label{prob:P2}
\begin{flalign}
\centering
 &\mathcal{P}_1: \max_{\substack{\mathbf{x}^p, \,\mathbf{V}, \,  \mathbf{w}_k , \,\mathbf{p}, \,\boldsymbol{\alpha},\boldsymbol{\beta}}} \quad  \text{SEE},\:\label{constm} \\
 &\text{s.t.} \quad ||\mathbf{w}_k||^2 = 1,\quad \forall k \in \mathcal{K},   \label{c1}\\
 &\qquad  R_k \geq R_{th}, \quad \forall k \in \mathcal{K}, \label{c2}\\
 & \qquad 0 \leq p_k \leq P_{k,max},\quad \forall k \in \mathcal{K}, \label{c3}\\
  & \qquad \mathrm{Tr}(\mathbf{V}) \le P_{BS}, \label{c4}\\
  & \qquad \mathbf{V} \succeq \mathbf{0}, \label{c9}\\
 & \qquad \alpha_m \in \{0,1\}, \quad \forall m \in \mathcal{M}, \label{c5}\\
 & \qquad \beta_m \in \{0,1\}, \quad \forall m \in \mathcal{M}, \label{c6}\\
 & \qquad \alpha_m+\beta_m \leq 1, \quad \forall m \in \mathcal{M}, \label{c7}\\
  & \qquad \psi_m \in [0,L], \quad \forall m \in \mathcal{M}, \label{c8}  
\end{flalign}
\end{subequations}
where $\textbf{x}^p=[\psi_1,\dots,\psi_m]$ represents the antenna positions for the $m$-th waveguide. $\textbf{p}=[p_1,\dots,p_k]$ represents for user power allocation. Constraint \eqref{c2} guarantees the minimum QoS requirements for the UL users, \eqref{c3} bounds the uplink transmit power, and \eqref{c4} limits the transmit power of the BS, and \eqref{c9} constraints positive semidefinite matrices, respectively. The conditions \eqref{c5} and \eqref{c6} impose that binary assignment indicator for segment selection and \eqref{c7} ensures that each segment can serve either uplink or downlink. Constraint \eqref{c8} ensures that the position of each PA is within the length of its segment. 

It can be seen that the formulated optimization problem is highly non-convex, due to the fractional and coupled objective function, the strong coupling among the optimization variables, and the presence of non-convex constraints.

\section{Solution Approach}\label{sec:algorithm}
In this section, the proposed SEE maximization problem is solved using a Dinkelbach-based AO framework. Specifically, the outer loop applies the Dinkelbach method to transform the fractional SEE objective into an equivalent subtractive form. For a given Dinkelbach parameter, the resulting problem is solved in the inner loop via AO by decomposing it into four subproblems: receive beamforming, joint user transmit power and AN optimization, antenna position update, and segment selection. The receive beamforming vector is obtained in a closed-form solution, while the joint transmit power and AN vectors are optimized using difference-of-convex (DC) programming. The antenna position variables are updated using SCA approach, in which first-order Taylor approximations are employed to convexify the non-convex terms at each iteration and the binary segment selection variables are handled via a logarithmic penalty method combined with SCA. The details of each subproblem are presented in the following subsections. 

\subsection{Subproblem 1: Receive Beamformer Design}

For given transmit powers $\mathbf{p}$, PA positions $\mathbf{x}^p$, and segment mode variables $\alpha$ and $\beta$, the receive beamformer $\mathbf{w_k}$ is only related to the UL achievable rate of user $k$ is given in \eqref{Rk_c}.

\allowdisplaybreaks
\begin{subequations}
\label{prob:P3}
\begin{flalign}
\centering
 &\mathcal{P}_2: \max_{\substack{  \mathbf{w}_k }} \quad  \text{SEE} \:\label{const2_1} \\
 &\text{s.t.} \quad \eqref{c1}, \eqref{c2}.
 \end{flalign}
\end{subequations}
Since $\log_2(1+x)$ is monotonically increasing in $x$, 
maximizing $R_k$ is equivalent to maximizing the uplink SINR in \eqref{eq:SINR_K}.
Define the interference-plus-noise covariance matrix for user $k$ as
\begin{equation}
\mathbf{A}_k
\triangleq
\sum_{i\neq k} p_i \mathbf{h}_i \mathbf{h}_i^{H}
+ \sigma_k^2 \mathbf{I}_{M},
\label{eq:Ak_def}
\end{equation}
where $\mathbf{I}_{M}$ denotes the $M \times M$ identity matrix.
Then, \eqref{eq:SINR_K} can be rewritten as the generalized Rayleigh quotient \cite{li2026movable}:
\begin{equation}
\gamma_{k}(\mathbf{w}_{k})
=
\frac{p_k \mathbf{w}_{k}^{H}
      \mathbf{h}_k \mathbf{h}_k^{H}
      \mathbf{w}_{k}}
{\mathbf{w}_{k}^{H}
 \mathbf{A}_k
 \mathbf{w}_{k}}.
\label{eq:SINR_matrix_form}
\end{equation}

Since the above function is a ratio of two quadratic forms 
with respect to $\mathbf{w}_{k}$, the optimal solution is obtained 
when $\mathbf{w}_{k}$ is aligned with the interference-whitened 
channel vector $\mathbf{A}_k^{-1}\mathbf{h}_k$. 
Imposing the unit-norm constraint $\|\mathbf{w}_{k}\|^2 = 1$, 
the closed-form solution is given by
\begin{equation}
\mathbf{w}_{k}^{\star}
=
\frac{\mathbf{A}_k^{-1}\mathbf{h}_k}
{\left\|\mathbf{A}_k^{-1}\mathbf{h}_k\right\|_2},
\quad \forall k=1,\ldots,K.
\label{eq:wr_closed_form}
\end{equation}

\subsection{Subproblem 2: Joint Power and Artificial Noise Design}

For given PA positions $\mathbf{x}^p$, duplex-mode variables 
${\alpha}$ and ${\beta}$, and receive beamformers $\mathbf{w}_k$, we optimize the user transmit powers 
$\mathbf{p}=[p_1,\dots,p_K]^T$ and the AN beamformer $\mathbf{V}$ while satisfying QoS and UE and BS power constraints. The corresponding subproblem is 

\begin{subequations}
\label{prob:P4}
\begin{equation}
\begin{aligned}
\mathcal{P}_3:\quad 
\max_{\mathbf{p},\mathbf{v}} \quad 
& \frac{R^{\mathrm{sec}}(\mathbf{p},\mathbf{V})}
{P_{\mathrm{tot}}(\mathbf{p},\mathbf{V})} \\
\text{s.t.}\quad 
& \eqref{c2}-\eqref{c9}.
\end{aligned}
\label{const4}
\end{equation}
\end{subequations}

where, the total secrecy rate is
\begin{equation}
R^{\mathrm{sec}}(\mathbf{p},\mathbf{V})
=
\sum_{k=1}^{K}R_k^{\mathrm{sec}}(\mathbf{p},\mathbf{V})
=
\sum_{k=1}^K\left[R_k^{}(\mathbf{p})
-
R_{j,k}^{}(\mathbf{p},\mathbf{V})
\right],
\end{equation}

and total power consumption is
\begin{equation}
P_{\mathrm{tot}}(\mathbf{p},\mathbf{V})
=
P_c
+
\xi \left(\sum_{k=1}^{K} p_k
+
\mathrm{Tr}(\mathbf V)\right).
\end{equation}

\subsubsection{BS-Side Achievable Rate}

Define the effective channel gains as $g_{i,k} \triangleq |\mathbf{h}_i^H \mathbf{w}_k|^2$, therefore, the uplink rate of user $k$ at the BS is given by
\begin{equation}
R_k^{}(\mathbf{p})
=
\log_2\!\left(
1+
\frac{p_k g_{i,k}}
{\sum_{i\neq k} p_i g_{i,k}
+ \sigma_k^2 \|\mathbf{w}_k\|^2}
\right),
\label{eq:R_k_BS}
\end{equation}

Let
\begin{align}
S_k^{}(\mathbf{p})
&\triangleq
\sum_{i=1}^{K} p_i g_{i,k}
+ \sigma_k^2 \|\mathbf{w}_k\|^2,\\
I_k^{}(\mathbf{p})
&\triangleq
\sum_{i\neq k} p_i g_{i,k}
+ \sigma_k^2 \|\mathbf{w}_k\|^2,
\end{align}

Then, the BS rate can be rewritten as the difference of two concave function forms as
\begin{equation}
R_k^{}(\mathbf{p})
=
\log_2\!\big(S_k^{}(\mathbf{p})\big)
-
\log_2\!\big(I_k^{}(\mathbf{p})\big).
\label{eq:RkB_DC}
\end{equation}

\subsubsection{Eve-Side Rate}

Define $a_{i,k} \triangleq |h_{i,j}|^2$ and $c(\mathbf{v}) \triangleq
\frac{1}{M_{\mathrm{DL}}}
|\mathbf h_{b,j}^{H}\mathbf V \mathbf h_{b,j}|$ and the eavesdropper rate is expressed as
\begin{equation}
\footnotesize
R_{j,k}^{}(\mathbf{p},\mathbf{v})
=
\log_2\!\left(
1+
\frac{p_k a_{i,j}}
{\sum_{i\neq k} p_i a_{i,j}
+ c(\mathbf{V})
+ \sigma_e^2}
\right),
\end{equation}

Similarly, we can define
\begin{align}
S_{j,k}^{}(\mathbf{p},\mathbf{V})
&\triangleq
\sum_{i=1}^K p_i a_{i,k}
+ c(\mathbf{V})
+ \sigma_e^2,\\
I_{j,k}^{}(\mathbf{p},\mathbf{V})
&\triangleq
\sum_{i\neq k} p_i a_{i,k}
+ c(\mathbf{V})
+ \sigma_e^2.
\end{align}

Thus,
\begin{equation}
R_{j,k}^{}(\mathbf{p},\mathbf{V})
=
\log_2\!\big(S_{j,k}^{}(\mathbf{p},\mathbf{V})\big)
-
\log_2\!\big(I_{j,k}^{}(\mathbf{p},\mathbf{V})\big).
\label{eq:RkE_DC}
\end{equation}

\subsubsection{Dinkelbach Reformulation for SEE}

Let $\eta \ge 0$ denote the auxiliary parameter representing the achievable SEE. 

According to Dinkelbach’s transformation \cite{zhang2020energy}, the original fractional problem is equivalent to solving the following parametric problem:
\begin{equation}
\max_{\mathbf{p},\mathbf{v}} 
F_\eta(\mathbf{p},\mathbf{v})
=
R^{\mathrm{sec}}(\mathbf{p},\mathbf{V})
-
\eta P_{\mathrm{tot}}(\mathbf{p},\mathbf{v}),
\end{equation}

At iteration $t$, the parameter $\eta$ is iteratively updated as
\begin{equation}
\eta^{(t+1)} =
\frac{R^{\mathrm{sec}}(\mathbf{p}^{(t)},\mathbf{V}^{(t)})}
{P_{\mathrm{tot}}(\mathbf{p}^{(t)},\mathbf{V}^{(t)})}.
\end{equation}

\subsubsection{DC Decomposition and SCA}

By grouping the logarithmic terms as in \cite{ding2024secure}, the secrecy rate can be written as
\begin{equation}
R^{\mathrm{sec}}(\mathbf{p},\mathbf{V})
=
\mathcal{P}(\mathbf{p},\mathbf{V})
-
\mathcal{Q}(\mathbf{p},\mathbf{V}),
\end{equation}
where
\begin{align}
\mathcal{P}(\mathbf{p},\mathbf{V})
&=
\sum_{k=1}^{K}
\log_2\!\big(S_k^{}\big)
+
\sum_{k=1}^{K}
\log_2\!\big(I_{j,k}^{}\big),\\
\mathcal{Q}(\mathbf{p},\mathbf{V})
&=
\sum_{k=1}^{K}
\log_2\!\big(I_k^{}\big)
+
\sum_{k=1}^{K}
\log_2\!\big(S_{j,k}^{}\big).
\end{align}

At SCA iteration $n$, $\mathcal{Q}(\mathbf{p},\mathbf{v})$ is upper-bounded by its first-order Taylor expansion around $(\mathbf{p}^{(n)},\mathbf{v}^{(n)})$:
\begin{equation}
\mathcal{Q}(\mathbf{p},\mathbf{V})
\le
\mathcal{Q}(\mathbf{p}^{(n)},\mathbf{V}^{(n)})
+
\nabla \mathcal{Q}^{(n)}
\big[(\mathbf{p},\mathbf{V})-(\mathbf{p}^{(n)},\mathbf{V}^{(n)})\big].
\end{equation}

Thus, a concave lower bound of secrecy rate is obtained as
\begin{equation}
\tilde{R}^{(sec,n)}(\mathbf{p},\mathbf{V})
=
\mathcal{P}(\mathbf{p},\mathbf{V})
-
\tilde{\mathcal{Q}}^{(n)}(\mathbf{p},\mathbf{V}).
\end{equation}

Accordingly, the concave surrogate of the Dinkelbach-transformed objective is expressed as
\begin{equation}
\tilde{F}_{\eta}^{(n)}(\mathbf{p},\mathbf{V})
=
\tilde{R}^{(\mathrm{sec},n)}(\mathbf{p},\mathbf{V})
-
\eta P_{\mathrm{tot}}(\mathbf{p},\mathbf{V}),
\label{eq:Feta_lb}
\end{equation}
which is solved to update $(\mathbf{p},\mathbf{V})$.

After solving the concave lower bound of the transformed objective, we next reformulate the QoS constraint of Subproblem 2. Although the minimum-rate requirement is non-convex in the original problem, it becomes an affine SINR constraint with respect to $\mathbf p$ when the receive beamformers are fixed. Hence, the $(\mathbf p,\mathbf V)$ update at each SCA iteration is performed by maximizing $\widetilde{F}_\eta^{(n)}(\mathbf p,\mathbf V)$ over a convex feasible set.

The minimum-rate requirement in (14b) is given by
\begin{equation}
    R_k \ge R_{\mathrm{th}}, \quad \forall k,
    \label{eq:Rth}
\end{equation}
Using the rate expression in \eqref{eq:R_k_BS}, the minimum rate constraint in \eqref{eq:Rth} can be equivalently transformed into the following:
\[
\gamma_{\mathrm{th}} \triangleq 2^{R_{\mathrm{th}}} - 1.
\]
Then $R_k \ge R_{\mathrm{th}}$ is equivalent to the SINR constraint
\begin{equation}
\frac{p_k | \mathbf h_k^{\mathrm H} \mathbf w_k |^2}{
\sum_{i \ne k} p_i | \mathbf h_i^{\mathrm H} \mathbf w_k |^2 
+ \|\mathbf w_k\|^2 \sigma_k^2}
\ge \gamma_{\mathrm{th}}.
\label{eq:constRth}
\end{equation}

After some mathematical simplifications, \eqref{eq:constRth} can be written
\begin{equation}
p_k | \mathbf h_k^{\mathrm H} \mathbf w_k |^2
\ge
\gamma_{\mathrm{th}}
\left(
\sum_{i \ne k} p_i | \mathbf h_i^{\mathrm H} \mathbf w_k |^2
+ \|\mathbf w_k\|^2 \sigma_k^2
\right),
\quad \forall k \in \mathcal{K}.
\label{eq:SINR_constraint_general}
\end{equation}
In the original problem $\mathcal P_1$, this constraint is non-convex since it is jointly dependent on $\mathbf p$ and $\mathbf w_k$. However, in the considered $(\mathbf p,\mathbf V)$ subproblem, the receive beamformers $\mathbf w_k$ are fixed and satisfy $\|\mathbf w_k\|^2 = 1$. Since the receive beamformer $\mathbf{w}_k$ is fixed in this subproblem, the effective channel gains $g_{i,k}$ become constants. Therefore,
\eqref{eq:SINR_constraint_general} reduces to
\begin{equation}
p_k g_{k,k}
\ge
\gamma_{\mathrm{th}}
\left(
\sum_{i \ne k} p_i g_{i,k}
+ \sigma_k^2
\right),
\quad \forall k \in \mathcal{K}.
\label{eq:SINR_linear_constraint_1}
\end{equation}

At iteration $n$, the $(\mathbf p,\mathbf V)$ update is obtained by solving

\begin{equation}
\max_{\mathbf p,\mathbf V}
\quad
\tilde F_\eta^{(n)}(\mathbf p,\mathbf V)
\end{equation}
\begin{align}
\text{s.t.}\quad
& p_k g_{k,k}
\ge
\gamma_{\text{th}}
\left(
\sum_{i\neq k} p_i g_{i,k}
+ \sigma_k^2
\right),
\quad \forall k, \label{eq:SINR_linear_constraint} \\
& \eqref{c3}-\eqref{c9}.
\end{align}

It can be observed that \eqref{eq:SINR_linear_constraint} is affine in the power vector $\mathbf p$, and thus, the QoS constraints (14b) define a convex feasible set in the $(\mathbf p,\mathbf V)$ subproblem. Consequently, in each Dinkelbach iteration, the SEE maximization subproblem is approximated as a convex problem via DC and SCA, and is efficiently solved using CVX with the MOSEK solver under constraints \eqref{eq:SINR_linear_constraint}, \eqref{c3}, and \eqref{c4}.

\subsection{Subproblem 3: PA Position Optimization}

Let $\mathbf{x}^p = [x_1,\ldots,x_M]^T$ denote the positions of the $M$ PAs along the waveguide. For given duplex-mode indicators ${\alpha}$ and ${\beta}$, user powers $\mathbf{p}$, AN beamformer $\mathbf{v}$, and receive beamformers $\mathbf{w}_k$, the effective channels
of user $k$ can be expressed as
\begin{equation}
\mathbf{h}_k = \mathbf{h}_k(\mathbf{x}^p), 
\qquad
h_{k,j} = h_{k,j}(\mathbf{x}^p),
\label{eq:hk_xp_def}
\end{equation}
where the explicit dependence on $\mathbf{x}^p$ comes from the 
free-space and in-waveguide propagation. Accordingly,
the BS and Eve rates in \eqref{eq:RkB_DC} and \eqref{eq:RkE_DC}
become functions of $\mathbf{x}^p$.

With all other variables fixed, the PA positioning subproblem aims at maximizing the total secrecy rate by optimizing $\mathbf{x}^p$.
The total secrecy rate is
\begin{equation}
R^{\mathrm{sec}}(\mathbf{x}^p)
=
\sum_{k=1}^{K}
R_k^{\mathrm{s}}(\mathbf{x}^p)
=
\sum_{k=1}^{K}
\Big(
R_k^{}(\mathbf{x}^p)
-
R_{j,k}^{}(\mathbf{x}^p)
\Big),
\end{equation}
where $R_k^{}(\mathbf{x}^p)$ and $R_{j,k}^{}(\mathbf{x}^p)$
are obtained by replacing the channels in \eqref{eq:RkB_DC} and 
\eqref{eq:RkE_DC} with \eqref{eq:hk_xp_def}.

Since the total power consumption does not depend on $\mathbf{x}^p$,
the SEE maximization w.r.t. $\mathbf{x}^p$ is equivalent to
maximizing the total secrecy rate. The PA positioning subproblem
is
\begin{equation}
\begin{aligned}
\max_{\mathbf{x}^p}\quad
& R^{\mathrm{sec}}(\mathbf{x}^p) \\
\text{s.t.}\quad
& \eqref{c8} 
\end{aligned}
\label{prob:xp_original}
\end{equation}
where \eqref{c8} ensure that all PAs lie within the waveguide length. Problem \eqref{prob:xp_original} is highly non-convex because
the secrecy rates $R_k^{}(\mathbf{x}^p)$ and
$R_{e,k}^{}(\mathbf{x}^p)$ depend on $\mathbf{x}^p$ through the channel magnitudes and phases. To tackle this,
we adopt an SCA approach. At iteration $n$, let $\mathbf{x}^{p,(n)}$ denote the current PA positions. For each
user $k$, we define
\begin{align}
S_k^{}(\mathbf{x}^p)
&\triangleq
\sum_{i=1}^{K} p_i 
\big|\mathbf{h}_i^H(\mathbf{x}^p)\mathbf{w}_k\big|^2
+ \sigma_k^2 \|\mathbf{w}_k\|^2, \\
I_k^{}(\mathbf{x}^p)
&\triangleq
\sum_{i\neq k} p_i 
\big|\mathbf{h}_i^H(\mathbf{x}^p)\mathbf{w}_k\big|^2
+ \sigma_k^2 \|\mathbf{w}_k\|^2, \\
S_{j,k}^{}(\mathbf{x}^p)
&\triangleq
\sum_{i=1}^{K} p_i |h_{i,j}(\mathbf{x}^p)|^2
+ c(\mathbf{v}) + \sigma_e^2, \\
I_{j,k}^{}(\mathbf{x}^p)
&\triangleq
\sum_{i\neq k} p_i |h_{i,j}(\mathbf{x}^p)|^2
+ c(\mathbf{v}) + \sigma_e^2.
\end{align}

The BS and Eve rates can then be written as
\begin{align}
R_k^{}(\mathbf{x}^p)
&=
\log_2 S_k^{}(\mathbf{x}^p)
-
\log_2 I_k^{}(\mathbf{x}^p), \\
R_{j,k}^{}(\mathbf{x}^p)
&=
\log_2 S_{j,k}^{}(\mathbf{x}^p)
-
\log_2 I_{j,k}^{}(\mathbf{x}^p).
\end{align}

Since the functions $S_k^{}(\mathbf{x}^p)$,
$I_k^{}(\mathbf{x}^p)$, $S_{e,k}^{}(\mathbf{x}^p)$
and $I_{e,k}^{}(\mathbf{x}^p)$ are differentiable with
respect to $\mathbf{x}^p$, we approximate them by their
first-order Taylor expansions around $\mathbf{x}^{p,(n)}$,i.e;
\begin{equation}
\begin{aligned}
S_k^{}(\mathbf{x}^p)
\approx\;
&\widetilde{S}_k^{(n)}(\mathbf{x}^p)
\triangleq
S_k^{}(\mathbf{x}^{p,(n)}) \\
&+
\nabla_{\mathbf{x}^p}
S_k^{}(\mathbf{x}^{p,(n)})^{T}
\big(\mathbf{x}^p-\mathbf{x}^{p,(n)}\big).
\end{aligned}
\end{equation}

Using these affine approximations, we obtain the following
SCA-based lower bound on the secrecy rate of user $k$:
\begin{equation}
\begin{aligned}
\widetilde{R}_k^{\mathrm{s},(n)}(\mathbf{x}^p)
=&\;
\log_2 \widetilde{S}_k^{(n)}(\mathbf{x}^p)
-
\log_2 \widetilde{I}_k^{(n)}(\mathbf{x}^p) \\
&-
\log_2 \widetilde{S}_{j,k}^{(n)}(\mathbf{x}^p)
+
\log_2 \widetilde{I}_{j,k}^{(n)}(\mathbf{x}^p).
\label{prob:xp_SCA}
\end{aligned}
\end{equation}

It can be seen that \eqref{prob:xp_SCA} a convex optimization problem because each $\widetilde{R}_k^{\mathrm{s},(n)}(\mathbf{x}^p)$ is expressed as a difference of logarithms of affine functions of $\mathbf{x}^p$, and the affine parts corresponding to the non-concave terms have been linearized via first-order Taylor expansion. Problem  \eqref{prob:xp_SCA} can be efficiently solved by standard convex solvers CVX.
\subsection*{Subproblem 4: Segment Selection Optimization}
Since the problem is mixed-integer and NP-hard, it is difficult to find the optimal solution; in order to find a low complexity solution, we relax the binary variables. We regularize the objective function with a penalty term and obtain an approximate solution with a regularization term \cite{xia2021joint} is defined as:
\allowdisplaybreaks
\begin{subequations}
\label{prob:P4}
\begin{flalign}
\centering
 &\mathcal{P}_2: \max_{\substack{ \,\boldsymbol\alpha,\boldsymbol\beta}} \quad  \text{SEE} - \rho f(\boldsymbol{\alpha},\boldsymbol{\beta})\:\label{const2} \\
 &\text{s.t.} \eqref{c5}- \eqref{c7}.
\end{flalign}
\end{subequations}

The sparsity inducing
penalty function is defined as:
\begin{equation}
\begin{aligned}
f(\boldsymbol{\alpha}, \boldsymbol{\beta})
= \sum_{m=1}^{M} \Big[
& \log(\alpha_m + \varepsilon)
+ \log(1 - \alpha_m + \varepsilon) \\
& + \log(\beta_m + \varepsilon)
+ \log(1 - \beta_m + \varepsilon)
\Big],
\end{aligned}
\label{eq:penalty}
\end{equation}
where $\varepsilon \geq 0$ and the logarithmic penalty function is used to smooth in order to obtain an exact penalty optimization approach for the original mixed-integer programming problem.

For fixed PA positions $\mathbf x^P$, AN vector $\mathbf v$, receive beamformers $\mathbf{w}_k$, and UL powers $\mathbf p$, the penalized problem $\mathcal P_2$ in \eqref{c3}--\eqref{c5} reduces to the following subproblem in the duplex-mode indicators
$\boldsymbol\alpha=[\alpha_1,\ldots,\alpha_M]^\top$ and
$\boldsymbol\beta=[\beta_1,\ldots,\beta_M]^\top$:
\begin{equation}
\begin{aligned}
\max_{\boldsymbol\alpha,\boldsymbol\beta}\quad
& F(\boldsymbol\alpha,\boldsymbol\beta) - \rho \, f(\boldsymbol\alpha,\boldsymbol\beta) \\
\text{s.t.}\quad
& \eqref{c5}-\eqref{c7}.
\end{aligned}
\label{eq:SP4_original}
\end{equation}
$F(\boldsymbol\alpha,\boldsymbol\beta)$ denotes the Dinkelbach-transformed SEE objective
\begin{equation}
   F(\boldsymbol\alpha,\boldsymbol\beta)
   \triangleq R^{sec}(\boldsymbol\alpha,\boldsymbol\beta)
   - \eta\, P_{\text{tot}}(\boldsymbol\alpha,\boldsymbol\beta),
\end{equation}
and the sparsity-inducing penalty $f(\boldsymbol\alpha,\boldsymbol\beta)$ is given in \eqref{eq:penalty} as with a small smoothing parameter $\varepsilon\ge 0$. The continuous relaxation $0\le\alpha_m,\beta_m\le 1$ together with the logarithmic penalty yields an exact penalty formulation that promotes $\alpha_m, \beta_m$ to be close to $0$ or $1$ \cite{xia2021joint}.

The problem in \eqref{eq:SP4_original} is still non-convex due to the non-concave SEE term $F(\boldsymbol\alpha,\boldsymbol\beta)$ and the penalty $f(\boldsymbol\alpha,\boldsymbol\beta)$. To obtain a tractable update, we adopt an SCA-based linearization around the current iterate
$(\boldsymbol\alpha^{(n)},\boldsymbol\beta^{(n)})$. Using the first-order Taylor expansion, we construct affine surrogates
{\footnotesize
\begin{align}
   \tilde F^{(n)}(\boldsymbol\alpha,\boldsymbol\beta)
   &= F^{(n)}
    + (\nabla_\alpha F^{(n)})^{\!\top}(\boldsymbol\alpha-\boldsymbol\alpha^{(n)})
    + (\nabla_\beta F^{(n)})^{\!\top}(\boldsymbol\beta-\boldsymbol\beta^{(n)}),
   \label{eq:F_tilde}\\
   \tilde f^{(n)}(\boldsymbol\alpha,\boldsymbol\beta)
   &= f^{(n)}
    + (\nabla_\alpha f^{(n)})^{\!\top}(\boldsymbol\alpha-\boldsymbol\alpha^{(n)})
    + (\nabla_\beta f^{(n)})^{\!\top}(\boldsymbol\beta-\boldsymbol\beta^{(n)}).
   \label{eq:f_tilde}
\end{align}
}
At iteration $n$, the $(\boldsymbol\alpha,\boldsymbol\beta)$-update is obtained by solving the following convex optimization problem:

\begin{equation}
\begin{aligned}
\max_{\boldsymbol\alpha,\boldsymbol\beta}\quad
 & \tilde F^{(n)}(\boldsymbol\alpha,\boldsymbol\beta)
   - \rho\, \tilde f^{(n)}(\boldsymbol\alpha,\boldsymbol\beta) \\
 \text{s.t.}\quad
 & \eqref{c5}-\eqref{c7}.
\end{aligned}
\label{prob:segment_SCA}
\end{equation}
The relaxed variables $(\boldsymbol{\alpha}, \boldsymbol{\beta})$ are updated based on the SEE gradient, which reflects the benefit of assigning each segment to uplink or downlink operation. For each segment, uplink reception (larger $\alpha_m$) or AN transmission (larger $\beta_m$) is selected depending on which configuration yields a higher SEE improvement. Since the objective in \eqref{prob:segment_SCA} is affine and all constraints are linear, \eqref{prob:segment_SCA} is a convex problem that can be efficiently solved using CVX solvers. The resulting solution $(\boldsymbol\alpha^{(n+1)},\boldsymbol\beta^{(n+1)})$ is then used as the starting point for the next AO–SCA iteration.

\begin{algorithm}[t]
\caption{AO--SCA--Dinkelbach Algorithm for SEE Maximization}
\label{alg:SEE}
\begin{algorithmic}[1]

\STATE Initialize $\mathbf{w}^{(0)},\mathbf{p}^{(0)},
\mathbf{V}^{(0)},\boldsymbol{\psi}^{(0)},
\boldsymbol{\alpha}^{(0)},\boldsymbol{\beta}^{(0)}$
and $\eta^{(0)}>0$. Set $t\gets 0$.

\STATE \textbf{Outer Dinkelbach loop:}
\REPEAT

    \STATE Set $n\gets 0$ and initialize
    $(\mathbf{w}^{(0)},\mathbf{p}^{(0)},
    \mathbf{V}^{(0)},\boldsymbol{\psi}^{(0)},
    \boldsymbol{\alpha}^{(0)},\boldsymbol{\beta}^{(0)})$.

    \STATE \textbf{Inner AO--SCA loop:}
    \REPEAT

        \STATE \textbf{(Subproblem 1)}
        Update $\mathbf{w}^{(n+1)}$ in closed form
        with fixed $\mathbf{p}^{(n)},\mathbf{V}^{(n)},
        \boldsymbol{\psi}^{(n)},
        \boldsymbol{\alpha}^{(n)},
        \boldsymbol{\beta}^{(n)}$.

        \STATE \textbf{(Subproblem 2)}
        Update $\mathbf{p}^{(n+1)},\mathbf{V}^{(n+1)}$
        by solving the convex SCA problem
        corresponding to \eqref{eq:Feta_lb}
        with parameter $\eta^{(t)}$.

        \STATE \textbf{(Subproblem 3)}
        Update $\boldsymbol{\psi}^{(n+1)}$
        by solving the convex SCA problem
        \eqref{prob:xp_SCA}.

        \STATE \textbf{(Subproblem 4)}
        Update $\boldsymbol{\alpha}^{(n+1)},
        \boldsymbol{\beta}^{(n+1)}$
        by solving the penalized objective
        \eqref{prob:segment_SCA}.

        \STATE Set $n\gets n+1$.

    \UNTIL{convergence of SEE for given $\eta^{(t)}$}

    \STATE Update the Dinkelbach parameter:
    \STATE $\displaystyle
    \eta^{(t+1)}\gets
    \frac{
    R^{\mathrm{sec}}(
    \mathbf{w}^{(n)},\mathbf{p}^{(n)},
    \mathbf{V}^{(n)},\boldsymbol{\psi}^{(n)},
    \boldsymbol{\alpha}^{(n)},\boldsymbol{\beta}^{(n)})
    }{
    P_{\mathrm{tot}}(
    \mathbf{p}^{(n)},\mathbf{V}^{(n)})
    }$.

    \STATE Set $t\gets t+1$.

\UNTIL{$\left|
R^{\mathrm{sec}}-\eta^{(t)}P_{\mathrm{tot}}
\right|\leq\delta$}

\STATE Threshold $\boldsymbol{\alpha},
\boldsymbol{\beta}$ to binary values.

\STATE \textbf{return}
$(\mathbf{w}^{\star},\mathbf{p}^{\star},
\mathbf{V}^{\star},\boldsymbol{\psi}^{\star},
\boldsymbol{\alpha}^{\star},
\boldsymbol{\beta}^{\star})$.

\end{algorithmic}
\end{algorithm}

\subsubsection{Convergence \& Complexity}

The proposed AO-SCA-Dinkelbach algorithm consists of an outer Dinkelbach loop and an inner AO-SCA loop. In each inner iteration, all variable blocks are updated either in closed form or by solving convex SCA approximations, ensuring that the objective value is non-decreasing. Since the SEE is upper bounded due to power constraints, the inner loop is guaranteed to converge. Moreover, the Dinkelbach method guarantees convergence of the outer loop \cite{dinkelbach1967nonlinear}. Therefore, the overall algorithm converges to a stable stationary solution.

Let $I_{\mathrm{out}}$ and $I_{\mathrm{in}}$ denote the numbers of outer Dinkelbach iterations and inner AO-SCA iterations, respectively. In each inner iteration, Subproblem~1 updates the receive beamforming vector $\mathbf{w}$ in closed form with negligible computational cost, while Subproblems~2, 3, and 4 are solved via convex optimization, whose computational complexities depend on the number of optimization variables. Using interior-point methods, the complexity of these subproblems is on the order of $\mathcal{O}(N_2^{3.5})$, $\mathcal{O}(N_3^{3.5})$, and $\mathcal{O}(N_4^{3.5})$ respectively \cite{boyd2004convex}. Hence, the total computational complexity of the proposed algorithm is
\begin{equation}
\mathcal{O}\!\left(
I_{\mathrm{out}} I_{\mathrm{in}}
(\mathcal{O}(N_2^{3.5})+\mathcal{O}(N_3^{3.5})+\mathcal{O}(N_4^{3.5}))
\right).
\end{equation}
Since the number of optimization variables is dominated by the number of segments $M$, the overall complexity is dominated by the segment dimension $M$.

\section{Simulation Results \& Discussions}
\label{sec:results}

We validate the proposed scheme through numerical simulations. We consider a rectangular service area of $D_x \times D_y = 100 \times 40$ m$^2$, in which the UEs and the eavesdropper are uniformly distributed. A strong LoS link is assumed between the UEs and the PASS. Unless otherwise stated, the system parameters are set as follows. The carrier frequency is $f_c = 2.8$ GHz, the effective refractive index of the waveguide is $n_{\text{eff}} = 1.4$, and the in-waveguide attenuation factor is $\kappa = 0.08$ dB/m. The wavelength is given by $\lambda = c/f_c$. The waveguide is deployed at a fixed height of 3 m and divided into $M=10$ segments of equal length.

We consider an uplink scenario consisting of multiple users and multiple eavesdroppers, and a minimum rate requirement for each user $R_{\text{th}} = 0.5$ bit/sec/Hz \cite{zeng2025energy}. The noise power at both the BS and Eve is $\sigma^2 = -90$ dBm. The maximum transmit power of each UE is $P_{k,\max}=10$ dBm, while the fixed circuit power consumption is set to 20 dBm \cite{ding2024secure}. To ensure statistical reliability, all results are averaged over 200 independent channel realizations. The proposed algorithm is initialized with feasible values for transmit power, beamforming vectors, antenna positions, and segment-selection variables. To further demonstrate the effectiveness of the proposed AO--SCA--Dinkelbach framework, we compare it with the following benchmark schemes:

\begin{itemize}
    \item \textbf{Segmented waveguide (SegWG) without AN}: No artificial noise is transmitted to interfere with the eavesdropper.
    \item \textbf{Fixed antenna position}: The PA position within each segment is fixed at the center and is not optimized.
    \item \textbf{Conventional PASS system (LongWG)}: A PASS architecture with a single long waveguide and multiple active antennas, where the IAR effect is neglected~\cite{jiang2026exploiting}. Two parallel long waveguides are deployed separately, one for uplink user reception and the other is dedicated to downlink AN transmission, respectively.
\end{itemize}

\begin{figure}
    \centering
\includegraphics[width=0.9\columnwidth]{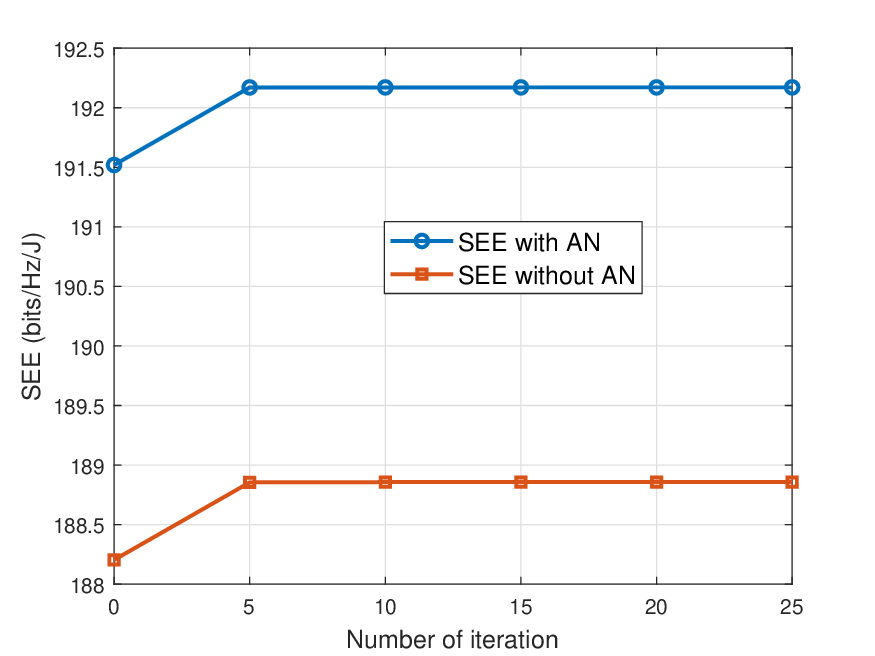}
    \caption{Convergence behavior of Algorithm 1 with $P_{\mathrm{BS}}=20$dBm and $P_{\mathrm{UE}}=20$dBm}
    \label{fig:convergence}
\end{figure}

Figure~\ref{fig:convergence} shows the convergence behavior of the proposed AO-SCA-Dinkelbach algorithm. The results indicate that the algorithm converges in approximately six iterations for both the AN-assisted and non-AN cases, confirming the fast convergence of the alternating optimization framework and its low computational complexity.


\begin{figure}
    \centering
\includegraphics[width=0.9\columnwidth]{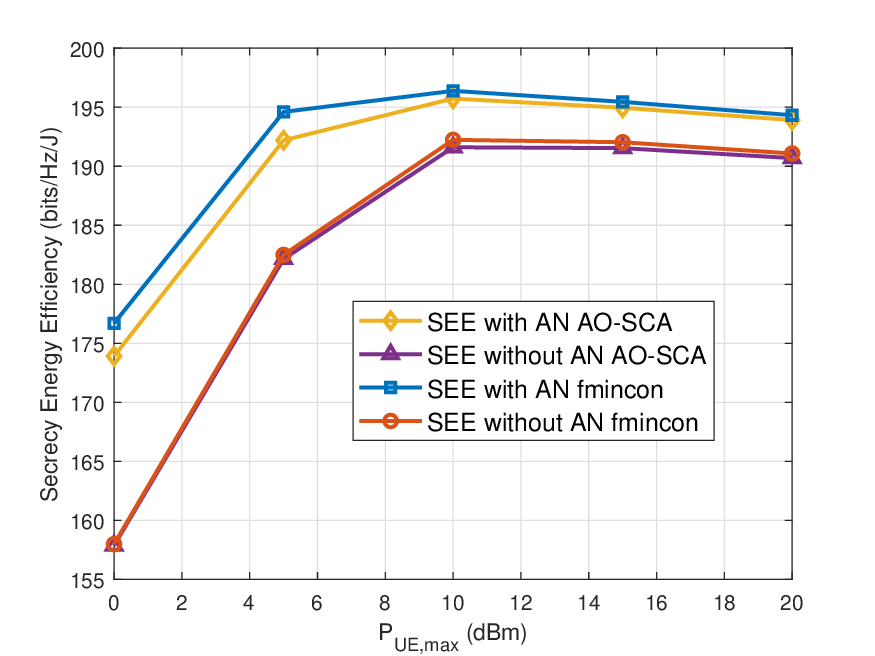}
    \caption{Secrecy Energy Efficiency vs User power for $K=2$, $J=1$}
    \label{fig:UEpower}
\end{figure}

Fig. \ref{fig:UEpower} presents the SEE as we vary the maximum transmit power of UEs, and compares the SEE performance obtained by the proposed AO-SCA algorithm and the benchmark fmincon solution (an off-the-shelf optimization solver). 200 multiple initial points are considered to improve the solver’s ability to converge to a near-optimal solution. It is observed that AO-SCA achieves near optimal performance to fmincon across all transmit power levels, validating the effectiveness of the proposed algorithm and the small performance gap is mainly due to successive convex approximations and alternating updates. The figure shows that as the UE power increases from 0 dBm to 10 dBm, the SEE improves for all schemes. This is because a moderate increase in UE transmit power substantially enhances the legitimate uplink secrecy rate while causing only a limited increase in power consumption, resulting in a significant SEE gain at low power levels. However, beyond approximately 10 dBm, the SEE begins to saturate because increasing the transmit power strengthens both the legitimate channels and the eavesdropper’s channel, thereby limiting the net secrecy gain. As a result, the marginal improvement in secrecy rate no longer offsets the additional transmit power consumption. In addition, the AN-assisted scheme consistently outperforms the non-AN case across all power levels, confirming the effectiveness of injecting AN through the downlink segments to suppress the eavesdropper’s reception. Specifically, at 0 dBm, the AN-assisted scheme achieves approximately 13\% higher SEE than the non-AN case, whereas this gain narrows to about 5\% near the saturation region at 10 dBm. This suggests that at higher UE power levels, the legitimate signal becomes sufficiently strong, reducing the relative contribution of AN to secrecy enhancement.

\begin{figure}
    \centering
\includegraphics[width=0.9\columnwidth]{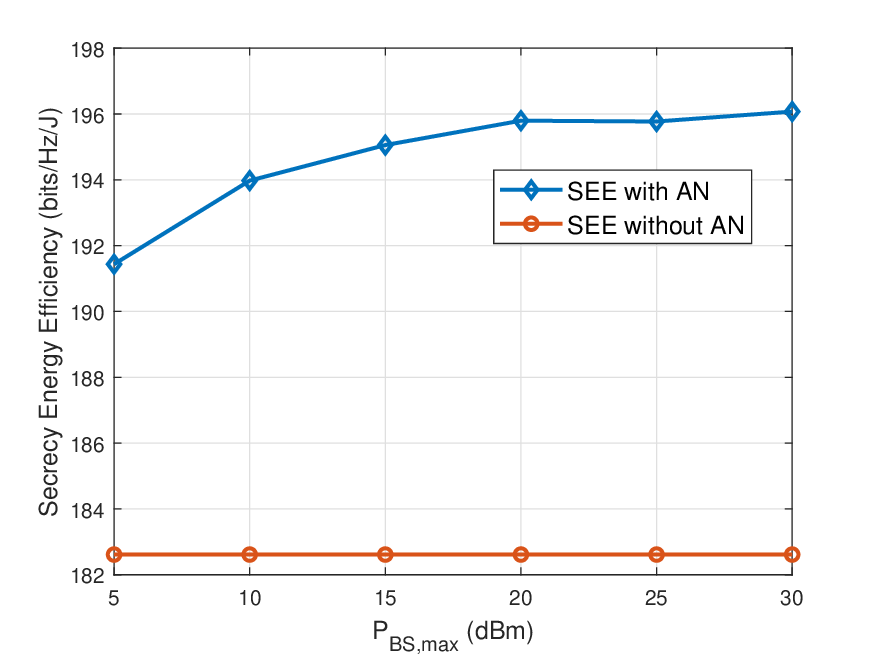}
    \caption{Secrecy Energy Efficiency Vs BS power for $K=2$, $J=1$}
    \label{fig:PBSpower_K2}
\end{figure}

\begin{figure}
    \centering
\includegraphics[width=0.9\columnwidth]{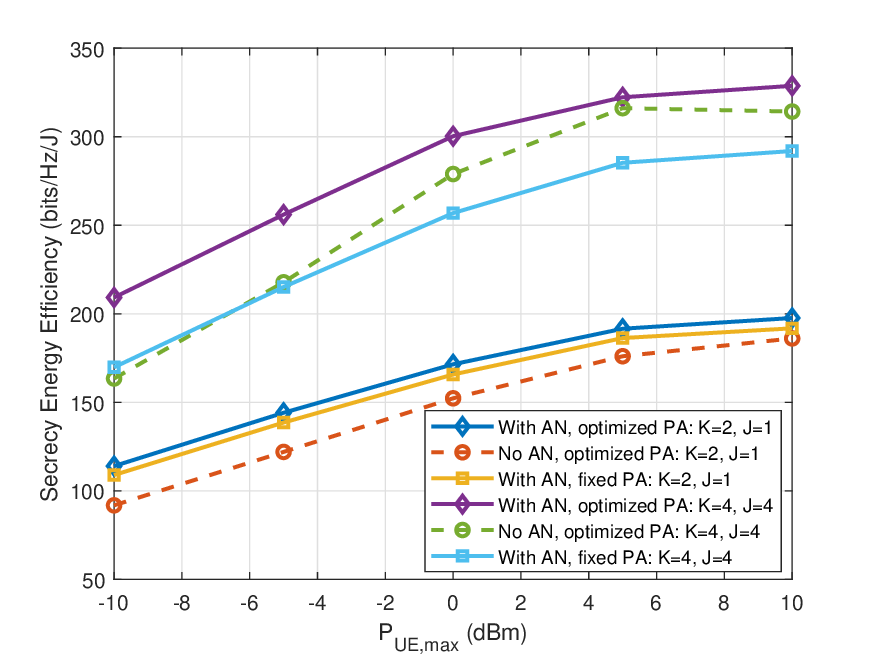}
    \caption{Comparison of SEE with fixed position antenna}
    \label{fig:fixedatcenter}
\end{figure}

Figure~\ref{fig:PBSpower_K2} depicts the SEE as we vary $P_{\mathrm{BS},\mathrm{max}}$. As expected, the scheme without AN shows less sensitive to $P_{\mathrm{BS}}$, since BS power does not directly affect secrecy performance in the absence of AN transmission. In contrast, when AN is employed, the SEE increases with $P_{\mathrm{BS},\mathrm{max}}$ in the low-power region, indicating that the additional BS power is effectively used to jam the eavesdropper’s received signal and prevent it from intercepting the uplink transmissions. Moreover, the SEE gain gradually saturates as the BS power increases, suggesting that once the eavesdropper’s channel is sufficiently degraded, further AN injection provides only limited secrecy improvement. At the same time, the additional AN transmission increases total power consumption, so the marginal secrecy gain no longer compensates for the extra energy cost.

\begin{figure}
    \centering
\includegraphics[width=0.9\columnwidth]{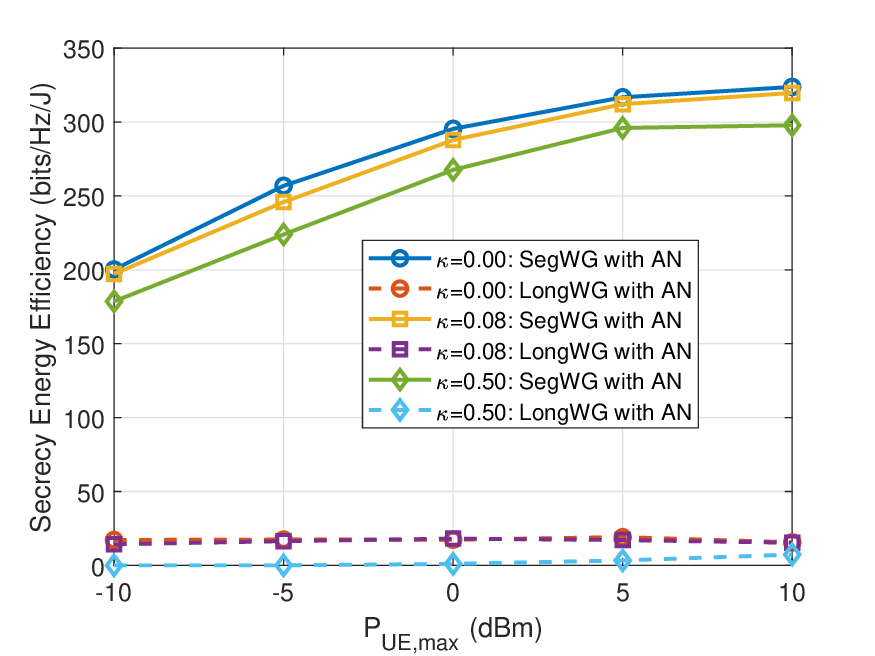}
    \caption{SEE comparison of uplink SWAN and conventional
PASS with different kappa value for $K=4$, $J=4$}
    \label{fig:multikappa}
\end{figure}

Figure~\ref{fig:fixedatcenter} illustrates the impact of PA flexibility and AN on SEE performance for both the $K=2$, $J=1$ and $K=4$, $J=4$ scenarios. In both cases, the movable PA scheme with AN achieves the best performance across all transmit power levels. Compared with the fixed-PA configuration, optimizing the antenna positions yields a consistent SEE improvement, as movable PAs can adjust their locations to establish more favorable uplink channels for legitimate users while simultaneously reducing information leakage to the eavesdropper. For the $K=2$, $J=1$ case, at 4 dBm, the movable-PA scheme improves SEE by approximately 9\% over the fixed-PA case, demonstrating the benefit of spatial adaptability.  In addition, the fixed-PA scheme with AN consistently outperforms the movable-PA scheme without AN, clearly confirming that AN plays a more dominant role in secrecy enhancement in the single-Eve scenario. In contrast, for the $K=4$, $J=4$ case, although the secrecy gain of AN becomes less pronounced due to the increased number of eavesdroppers, and the difficulty of simultaneously suppressing multiple eavesdropping links, the results clearly demonstrate the benefits of antenna position optimization. Despite the presence of more Eves, $K=4$ and $J=4$ scenario achieves significantly higher SEE because the increased number of users provides higher multiplexing gain and spatial diversity, which improves the overall achievable secrecy sum rate. Specifically, optimizing PA locations provides substantial improvements in secrecy performance by enhancing the legitimate users’ channels and exploiting spatial diversity more effectively. Furthermore, the AN-assisted schemes significantly outperform the non-AN case at all power levels, especially in the low-power regime, where the legitimate signal strength is limited and AN is effective in suppressing the eavesdropper’s rate. As the transmit power increases, the performance gap gradually decreases because both the legitimate users and the eavesdropper receive stronger signals, reducing the relative impact of AN and antenna positioning on improving the secrecy rate. The results demonstrate that AN and PA mobility jointly improve SEE performance, while antenna position optimization plays a crucial role in complex multi-eavesdropper environments.

\begin{figure}
    \centering
\includegraphics[width=0.9\columnwidth]{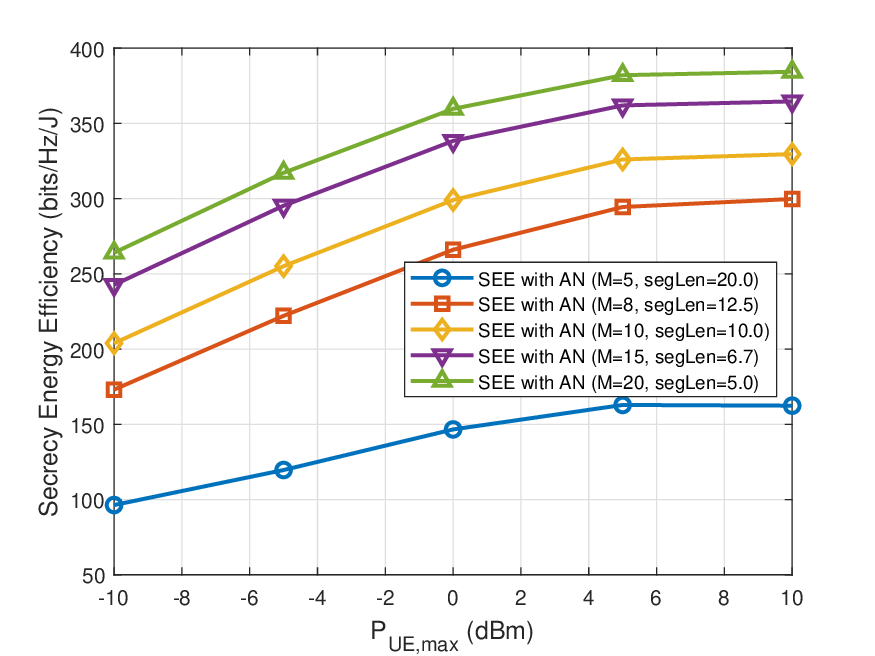}
    \caption{Secrecy Energy Efficiency vs Number of Segments for $K=4$, $J=4$}
    \label{fig:multiuserseg}
\end{figure}

Fig.\ref{fig:multikappa} examines the SEE versus UE transmit power for different in-waveguide attenuation factors $\kappa$ comparing the segmented waveguide (SegWG) and conventional LongWG schemes for the $K=4$, $J=4$ scenario. It is observed that the SEE decreases as $\kappa$ increases for both architectures, indicating that larger attenuation factors introduce higher propagation loss inside the waveguide, thereby reducing the effective channel gain and SEE performance. Moreover, the proposed SegWG architecture consistently outperforms the LongWG scheme across all values of $\kappa$. This performance gap becomes even more pronounced at higher attenuation. This behavior occurs because the LongWG architecture suffers from accumulated in-waveguide propagation loss over the entire waveguide length, whereas the segmented structure reduces the average propagation distance between the feed point and the PAs, thereby significantly mitigating attenuation effects. Moreover, the SegWG scheme shows better robustness to attenuation. This behavior aligns with the structural advantage of segmentation: each segment has an independent feed point connected directly to the BS, which mitigates propagation loss by shortening the signal transmission distance. Finally, an optimal UE transmit power is observed around 10 dBm, beyond which the SEE slightly saturates, since the secrecy-rate improvement becomes marginal at high transmit-power levels while the total power consumption continues increasing.

Fig.\ref{fig:multiuserseg} shows the SEE performance for different numbers of segments under a fixed total waveguide length $L=100$m for $K=4$ and $J=4$. It is observed that increasing the number of segments (i.e increasing the number of antennas) significantly improves SEE, due to enhanced spatial flexibility, higher beamforming capabilities and reduced in-waveguide propagation loss within each segment. Specifically, dividing the total waveguide into smaller segments shortens the average propagation distance between the feed point and the PAs, thereby mitigating attenuation and improving the effective channel gain. Moreover, finer segmentation enables more effective channel shaping for legitimate users while reducing information leakage to eavesdroppers. The gain is particularly significant when increasing the number of segments from 5 to 10, and further increasing to 20 and 25 yields 5–12\% additional gain. In particular, the smaller segmentation (e.g.M=5) exhibits substantially lower performance, highlighting the importance of finer segmentation in this setting since longer segment length results in more severe accumulated in-waveguide attenuation and limited antenna-position flexibility. However, the improvement becomes marginal as the number of segments increases beyond a certain point, showing a clear saturation effect. 



\begin{figure}[!t]
    \centering
\includegraphics[width=0.9\columnwidth]{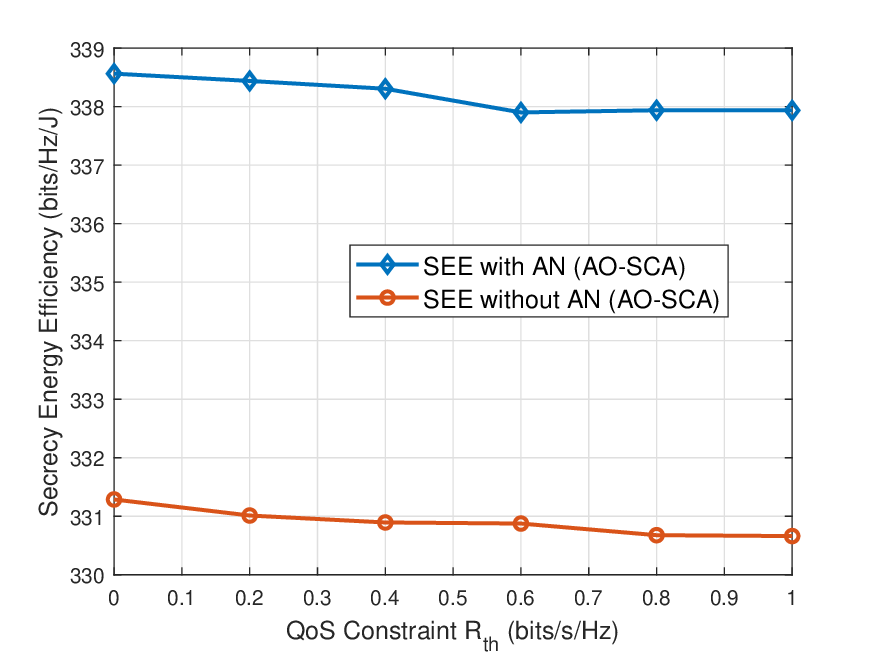}
    \caption{Multiuser rate threshold vs SEE for $K=4$, $J=4$}
    \label{fig:multiuserrateth}
\end{figure}

Fig.~\ref{fig:multiuserrateth} shows the impact of rate threshold $R_{th}$ on SEE for $K=4$, $J=4$ scenario, comparing with or without AN. It can be observed that the SEE gradually decreases as the rate threshold $R_{th}$ increases for both schemes. This behavior is expected because enforcing a higher minimum rate constraint requires the transmitter to allocate more power to satisfy minimum rate constraint $R_{th}$ for each user, which reduces the overall energy efficiency. Moreover, the proposed scheme with AN consistently achieves higher SEE compared to no AN scheme for all $R_{th}$ values, with a relatively stable performance gap, indicating that AN provides a consistent secrecy improvement even under stricter rate constraints.

\begin{figure}
    \centering
\includegraphics[width=0.9\columnwidth]{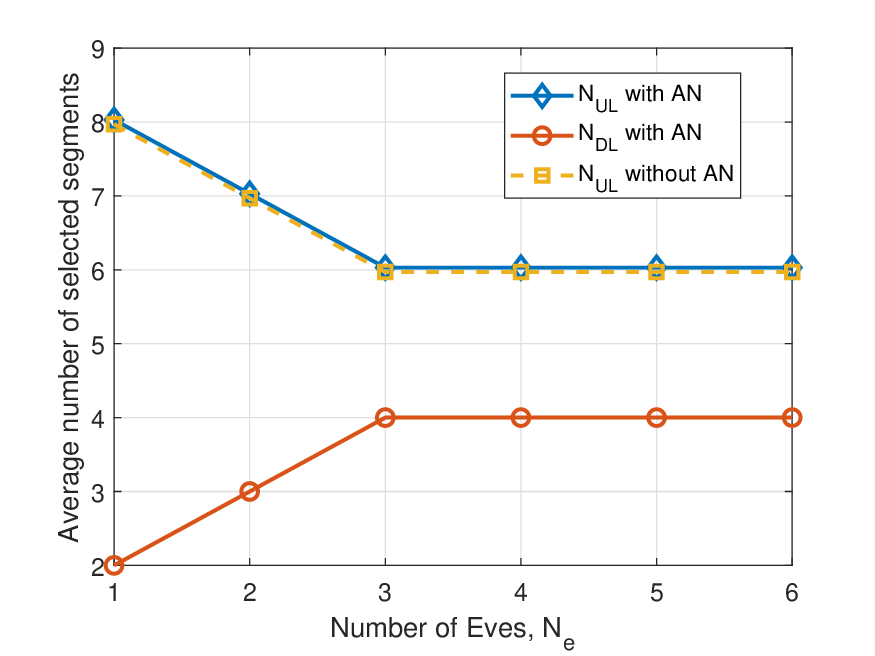}
    \caption{Uplink/Downlink Segment Selection vs. Number of Eves}
    \label{fig:multiEveULDL}
\end{figure}

Fig.~\ref{fig:multiEveULDL} illustrates the average number of selected UL and DL segments as the number of Eves increases. This behavior occurs because the presence of more Eves increases the risk of eavesdropping and requires the system to secure a large spatial region. The optimizer allocates more segments to the DL/AN mode in order to improve AN coverage and suppress multiple eavesdropping links simultaneously. However, after $N_e=3$, both UL and DL segment allocations become stable, indicating that the existing AN resources are already sufficient to provide effective jamming for additional Eves, without allocating additional segments from the UL mode. Moreover, the UL segment allocation in the noAN case closely follows that of the AN-assisted scheme. This is because, for a fair comparison, the noAN scheme follows the same segment selection strategy as the AN-assisted case. Hence, the decrease in UL segments is mainly caused by the increased secrecy requirement due to the larger number of Eves, rather than by the AN transmission itself. Overall, the figure clearly shows the tradeoff between allocating more DL segments to improve secrecy performance for multiple eves and maintaining enough UL segments for efficient multiuser uplink communication.

\section{Conclusion}
\label{sec:conclusion}

This paper proposed a segmented waveguide-enabled uplink PASS with artificial noise to maximize SEE for improved physical layer security. The SEE maximization problem was formulated and solved using an AO-SCA-Dinkelbach framework. Numerical results showed that the proposed approach achieves near-optimal performance and significantly outperforms conventional long waveguide systems in terms of SEE and robustness against waveguide attenuation. The results demonstrated that AN effectively suppresses information leakage, while antenna mobility and waveguide segmentation further improve SEE through flexible PA positioning and reduced in-waveguide propagation loss. The results also illustrated the impact of waveguide segmentation, where increasing the number of segments improves SEE by reducing propagation loss and enhancing spatial flexibility. Future work will extend the framework to a joint uplink-downlink SEE optimization incorporating simultaneous AN transmission and information reception, particularly given the inherent full-duplex capability of the PASS architecture and \textit{robust secrecy design under imperfect CSI}. In addition, integrating AI-assisted dynamic segment allocation represents another promising research direction for improving scalability and adaptability in practical deployments.
 
\bibliographystyle{IEEEtran}
\bibliography{references}
\vfill
\end{document}